\documentclass{aa}  

\usepackage{graphicx}
\usepackage{gensymb}
\usepackage{natbib}
\usepackage{diagbox}  
\usepackage{siunitx}
\bibpunct{(}{)}{;}{a}{}{,}
\usepackage[colorlinks=true, allcolors=blue]{hyperref}
\usepackage{txfonts}

\titlerunning{Spatial distribution of star formation in dwarf galaxies: Merger-driven and secular starburst}
\authorrunning{Chauhan et al.}
\begin{document} 

  \title{Spatially Extended Star Formation in Merger-Driven and Secular Starbursts in Dwarf Galaxies}

   \author{ Rakshit Chauhan
          \inst{1,2},
          Smitha Subramanian\inst{1,2,3}
          }

   \institute{Indian Institute of Astrophysics, Koramangala II Block, Bangalore-560034, India \\
              \email{rakshit.chauhan@iiap.res.in}
         \and
            Pondicherry University, R.V. Nagar, Kalapet, 605014, Puducherry, India 
            \and
            Leibniz-Institut für Astrophysik Potsdam (AIP), An der Sternwarte 16, D-14482 Potsdam, Germany}

  \date{Received XXXX; accepted XXXX}

  \abstract
   {Galaxy interactions can induce bursts of star formation, while internal or secular processes can also trigger starbursts. Recent studies of the spatial distribution of star formation in high-mass (M$_{*}$ $>$ 10$^{10}$ M$_{\odot}$) starburst galaxies show that star formation is centrally concentrated in merger-driven starbursts, whereas in secularly driven starburst systems, star formation is uniformly enhanced across the system. However, it is not clear whether such distinct spatial signatures are present in low-mass (M$_{*}$ $<$10$^{10}$M$_{\odot}$) starburst galaxies. In this context, we studied the star formation properties of a sample of 272 isolated starburst (i.e., with no massive galaxy, M$_{*}$ $>$ 10$^{10}$ M$_{\odot}$, within a 1 Mpc$^3$ volume) dwarf galaxies consisting of 17 post-merger and 255 non-interacting starburst galaxies, spanning a stellar mass range of $ 10^{7} - 10^{9.6} M_{\odot} $ and a redshift range of 0.01 -- 0.12. For this study, we utilise far ultraviolet imaging data from the GALEX mission to estimate the star formation rates of our sample galaxies. We investigated the spatial variation of starburst activity by comparing star formation rate surface density (SFRD) across radial bins for post-merger and non-interacting samples of dwarfs. We find an enhancement in SFRD across all radial bins for both samples, and that the enhancement is more or less uniform across the galaxy, within uncertainties. This indicates that star formation in low-mass starburst galaxies is uniformly enhanced throughout the galaxy, regardless of the mechanism that triggers the starburst. This contrasts with what is observed in massive galaxies, where star formation is centrally concentrated in the merger-driven starburst systems. The lack of central concentration of star formation in low-mass merger systems could be due to their shallow gravitational potentials and the compression of extended gaseous envelopes during interactions, which trigger star formation across these galaxies. Deeper, higher-spatial-resolution observations are essential for better understanding the spatial distribution of star formation in low-mass galaxies.}

   \keywords{galaxies: dwarf --
                galaxies: starburst --
                galaxies: interactions --
                galaxies: irregular --
                galaxies: evolution --
                galaxies: star formation 
               }

   \maketitle
%
\section{Introduction}
Starburst (SB) galaxies are systems undergoing short-lived episodes of intense star formation compared to their long-term average star formation rates, and gas depletion times much shorter than the Hubble time (\citealt{1972ApJ...173...25S,1973ApJ...179..427S,1998ARA&A..36..189K,2009ApJ...698.1437K}). In the local universe, starbursts create approximately 20\% of the massive stars \citep{1998ApJ...503..646H,2004MNRAS.351.1151B}.
Numerous studies have demonstrated that massive galaxy interactions can trigger SB or enhance star-formation, and it depends on factors such as separation between the interacting galaxies, mass ratio, and gas fraction  (e.g., \citealt{2008AJ....135.1877E,2010MNRAS.407.1514E,2011MNRAS.418.2043E,2013MNRAS.435.3627E,2011MNRAS.412..591P,2013MNRAS.433L..59P,2016MNRAS.461.2589P,2012MNRAS.426..549S,2015MNRAS.449.3719S,2014MNRAS.441.1297S,2015MNRAS.452..616D,2019ApJ...881..119P,Steffen_2021}). While mergers are linked to SBs, elevated star-formation rate (SFR) can also result from other factors, such as disk instabilities, which can lead to gas infall and a centrally concentrated burst of star formation \citep{2015MNRAS.449..820W,2016MNRAS.457.2790T,2017ApJ...838..105L}. Additionally, gas accretion from intergalactic filaments could also trigger intense star formation in the host galaxy \citep{2013seg..book....1K,2016ASSL..418..355B}. Stellar bars are also capable of triggering central starburst activity as their gravitational torque is highly effective at redistributing angular momentum within the galactic disk \citep{10.1093/mnras/259.2.328,2012MNRAS.423.3486W,1986MNRAS.221P..41H,1997A&A...323..363M,10.1111/j.1365-2966.2011.19195.x,2012ApJS..198....4O,2017ApJ...845...93K,10.1093/mnras/stz349}. 
Thus, SBs in galaxies could be triggered by merger events or by more gradual, secular processes. \\
\indent A key question is whether the resulting star formation, triggered by merger events or secular processes, exhibits distinct observable properties.
In this regard, \cite{2024A&A...690L...4T} undertook a detailed study of the spatial distribution of star formation in massive SB galaxies having stellar masses $\geq 10^{9.5} M_{\odot}$ using observations from the Mapping Nearby Galaxies at Apache Point Observatory (MaNGA) survey. Their investigation showed a clear distinction between merger-driven and non-merger-driven SB massive galaxy populations. Among the two sets of SB galaxies with comparable overall SFR enhancement, merger-driven SB galaxies were found to possess a much stronger centrally concentrated burst of star formation than isolated SB galaxies. In contrast, isolated SB galaxies showed a more uniform enhancement in the SFR surface density 
across different galactocentric distances. These findings suggest that galaxy mergers in the massive regime efficiently channel gas and trigger SBs predominantly in the galactic nucleus, whereas secular processes trigger uniform star formation across the galaxy. However, an important question arises: how do these processes operate in dwarf galaxies? 

\indent According to the hierarchical $\Lambda$ cold dark matter ($\Lambda$CDM) paradigm, dwarf galaxies are the most numerous and the first to form; therefore, the majority of mergers are expected to occur between them \citep{2008ApJ...683..597S,2010MNRAS.406.2267F,2015ApJ...807...49W}. Hence, SBs in interacting dwarf galaxies are expected.  The shallow potential wells of dwarf galaxies make them highly sensitive to perturbations such as gravitational interactions  \citep{2001ApJ...559..754M, 2006MNRAS.369.1021M}, ram-pressure stripping \citep{1972ApJ...176....1G,2003AJ....125.1926G,2008ApJ...674..742B}, and feedback from supernovae and stellar winds \citep{1986ApJ...303...39D,1999ApJ...513..142M,2010Natur.463..203G}. 
The impact of SBs may be a crucial factor in understanding the potential connection between the evolution of dwarf galaxies (e.g., \citealt{1986sfdg.conf...73L,1986ApJ...303...39D,1987IAUS..115..663S,1988MNRAS.232..239D}) with external effects, such as ram pressure and tidal stripping resulting from interactions with other systems (\citealt{2001ApJ...559..754M,2001ApJ...547L.123M,2004AJ....128.2797V,2006MNRAS.369.1021M}). The intensity of the feedback processes and their effects depend on the star-formation activity within these systems. Thus, SB activity in dwarf galaxies plays an important role in their evolution. 
However, it is unclear whether SBs are common to most dwarfs or restricted to a subgroup of dwarfs with specific intrinsic properties or unique environmental interactions \citep{1997ApJS..112..285M,2001ApJ...559..754M,2001AJ....122..121V,2003A&A...405..931P}.

With the technological advancements and the increasing availability of high-resolution, multiwavelength observational data, we can now study star formation in interacting dwarf galaxies across a broad range of stellar masses and redshifts (e.g; \citealt{2014A&A...566A..71L, 2015ApJ...805....2S, Privon_2017, 2024A&A...681A...8S,2024ApJ...963...37K, 2025A&A...702A.222C}). Many of these studies show that galaxy interactions can enhance star formation in the dwarf galaxy regime. However, some galaxies also exhibit signatures of quenching, indicating that galaxy interactions can also suppress star formation. We note that because of the selection biases included in shallow surveys, most of the studies in the literature may not represent the dwarf galaxy population as a whole \citep{2025MNRAS.538..153K}. While most of the sample of low-mass interacting galaxies studied in the literature exhibit moderate enhancement, some show SFRs 5–10 times higher than expected for their stellar mass, resulting in SB. This suggests that interactions can trigger starbursts in low-mass galaxies.
 Several studies related to individual SB dwarfs have found that the SB phenomenon is closely correlated with the intrinsic stellar and gaseous properties of their host galaxies \citep{2012A&A...537A..72L,2012A&A...544A.145L}. Starburst events in dwarfs may not be confined to the central region only, but they could cover a larger extent of the galaxy \citep{2012ApJ...759...77M, Privon_2017}. These episodes might be initiated by internal, secular processes such as torques induced by massive star-forming clumps \citep{2012ApJ...747..105E}, triaxial dark matter haloes \citep{2002ApJ...574L..21B}, or presence of dark matter bars \citep{2004AJ....128.2170H}, or solely by external factors \citep{2013AJ....146...46K,2021MNRAS.500.4937M} like tidal perturbations from nearby companions \citep{1988A&A...201...37N}, interactions/mergers between gas-rich dwarfs \citep{2008MNRAS.388L..10B}, or the accretion of cold gas from the intergalactic medium \citep{1987ApJ...322L..59S}.
However, whether the consequent changes in star-forming efficiency, spatial distribution, and kinematic characteristics of starburst regions triggered by external (interaction/merger-driven) events differ intrinsically from those induced by internal, or secular, events remains unexplored. In this context, we study a sample of merger-driven and non-merger-driven SB dwarf galaxies. 

Recently, \cite{2025A&A...702A.222C} (hereafter referred to as \href{https://www.aanda.org/articles/aa/full_html/2025/10/aa54437-25/aa54437-25.html}{Paper I}) studied a sample of dwarf galaxies (stellar mass from $10^{7}$ to $10^{9.6} M_{\odot}$) to understand the impact of dwarf-dwarf interactions on their star formation. The sample comprised 6155 spectroscopically confirmed (z $<$ 0.12) isolated dwarf galaxies (with no massive galaxy, M$_{*}$ $>$ 10$^{10}$ M$_{\odot}$, within a 1 Mpc$^3$ volume), with 194 dwarfs showing tidal features which could be the remnant of recent merger activity. 
These 6155 galaxies were selected based on the availability of Far Ultraviolet (FUV) data from the GALEX mission and the detectability of the sources in the images using the astrodendro algorithm (refer to Fig. 1 of \href{https://www.aanda.org/articles/aa/full_html/2025/10/aa54437-25/aa54437-25.html}{Paper I} for more details of the selection criteria), from a parent sample of 6875 galaxies given in the catalogue by \cite{2020AJ....159..103K}. \cite{2020AJ....159..103K}  used high resolution imaging data obtained as part of the Hyper Suprime-cam Subaru Strategic Program (HSC-SSP) and an automated tidal feature detection algorithm to identify the post-merger dwarf sample, with spectroscopic redshift from either GAMA or SDSS surveys.
In \href{https://www.aanda.org/articles/aa/full_html/2025/10/aa54437-25/aa54437-25.html}{Paper I}, we estimated the SFR of the sample galaxies using FUV imaging data from the GALEX mission.
Overall, a moderate enhancement (by a factor of nearly two) in the median SFR of the post-merger dwarf sample, compared to that of the non-interacting dwarfs, was found in \href{https://www.aanda.org/articles/aa/full_html/2025/10/aa54437-25/aa54437-25.html}{Paper I}. 
This factor is comparable to the average SFR enhancement factor observed in massive post-merger galaxies. However, star formation is found to be enhanced in both the central (6" diameter region at the centre) and outer regions (outside the central 6" diameter region) of post-merger star-forming galaxies compared to their non-interacting counterparts, matched in mass and redshift. This contrasts with what is observed in massive galaxies, where merger-triggered star formation is more significant in their central regions. 
As the aim of \href{https://www.aanda.org/articles/aa/full_html/2025/10/aa54437-25/aa54437-25.html}{Paper I} was to find the effect of mergers on star formation, all post-merger and non-interacting dwarf galaxies in the sample were included.
The majority of galaxies in the sample were star-forming, with some SB and quenched dwarfs. However, SB galaxies were not analysed separately. The present study specifically focuses on starburst dwarfs, including both post-merger and non-interacting systems.\\
\indent In this study, we use the SFR estimates and the star-forming main sequence (SFMS) relation provided in \href{https://www.aanda.org/articles/aa/full_html/2025/10/aa54437-25/aa54437-25.html}{Paper I}, to identify the SB sample among the post-merger dwarfs and non-interacting dwarfs. The SB in these two samples are expected to be triggered by distinct mechanisms: interaction-driven in post-merger dwarfs and secular processes driven in non-interacting dwarfs. The spatial distribution of star formation in both samples is analysed with respect to the SFMS, matched simultaneously in stellar mass and redshift. This will allow us to study whether SB activity triggered by the two different mechanisms exhibits distinct signatures in their spatial distribution, as observed in massive galaxies, or not.

The paper is structured as follows. The sample selection is presented in Section \ref{2}, the data and analysis are outlined in Section \ref{3}, and the results and discussion are presented in Sections \ref{4} \& \ref{5}. The summary of the study is presented in Section \ref{6}. Throughout this paper, we assume the standard $\Lambda$CDM cosmology model with parameters h = H$_0$/(100 km$s^{-1}Mpc^{-1}$) = 0.6774, Ω$_m$ = 0.3075, Ω$_ \Lambda$ = 0.691, Ω$_b$ = 0.0486 \citep{2016A&A...594A..13P}.

\section{Sample selection and data} \label{2} In this study, we analyse and compare the spatial variation of star formation in post-merger and non-interacting SB dwarfs.
The SB dwarfs are selected from the sample of 6155 (194 post-merger and 5961 non-interacting) dwarf galaxies, studied in \href{https://www.aanda.org/articles/aa/full_html/2025/10/aa54437-25/aa54437-25.html}{Paper I}. 
The FUV-based global SFR of these 6155 galaxies was estimated in \href{https://www.aanda.org/articles/aa/full_html/2025/10/aa54437-25/aa54437-25.html}{Paper I}, using the archival science-ready FUV imaging data from the GALEX mission, and by applying the calibration relation (connecting the FUV luminosity and SFR) given by \cite{Murphy_2011}. We note that the calibration constant used in the relation is for solar metallicity, and the observed flux is corrected only for Galactic extinction. However, dwarf galaxies generally have sub-solar metallicities and SFR calibration also shows a dependence on metallicity (metal-poor stellar populations can produce higher UV luminosities \citealt{2005A&A...443L..19B}), which could affect the absolute SFR values for our sample of SB dwarfs \citep{2010A&A...523A..64R,2012MNRAS.427.1490W}. This study involves a comparative analysis of star formation between isolated and post-merger SB dwarfs, matched in mass and redshift. Since the two subsamples of similar masses are expected to have similar metallicities, any systematic effect arising from the choice of SFR calibration is expected to influence both subsamples in a similar manner. Therefore, the choice of calibration is not expected to significantly affect our final results. Further details about the estimation of the global SFR of the sample galaxies can be found in \href{https://www.aanda.org/articles/aa/full_html/2025/10/aa54437-25/aa54437-25.html}{Paper I}. They used the stellar mass and redshift provided in the catalogue by \cite{2020AJ....159..103K}, which were based on the SDSS and GAMA spectroscopic surveys. In this study, we use these parameters, including the SFR estimated in \href{https://www.aanda.org/articles/aa/full_html/2025/10/aa54437-25/aa54437-25.html}{Paper I}.\\
\indent To select the SB dwarfs, we refer to Fig.5 of \href{https://www.aanda.org/articles/aa/full_html/2025/10/aa54437-25/aa54437-25.html}{Paper I}, which illustrates the position of all sample dwarfs on the SFR vs M$_{*}$ plot. 
Although in \href{https://www.aanda.org/articles/aa/full_html/2025/10/aa54437-25/aa54437-25.html}{Paper I}, separate linear fits were performed to the post-merger and
non-interacting dwarf galaxy samples, the best-fit parameters were similar (within errors). 
In this study, all those dwarfs that lie above the best-fit + RMS line (0.445 dex and 0.447 dex for post-merger and non-interacting dwarfs, respectively) are classified as SB dwarfs. The best-fit and RMS were determined by weighted least-squares (WLS) linear regression in \href{https://www.aanda.org/articles/aa/full_html/2025/10/aa54437-25/aa54437-25.html}{Paper I}, and the corresponding equations for the post-merger and non-interacting SBs are given as:

\begin{align}
     ~~log(SFR) = (0.709 \pm 0.055)~log(M_{*}) + (-7.724 \pm 0.500) \label{0.1} \\   
     ~~log(SFR) = (0.651 \pm 0.012)~log(M_{*}) + (-7.222 \pm 0.114) \label{0.2} 
\end{align}
\begin{figure}[ht!]
   \centering
   \includegraphics[width=9.1cm]{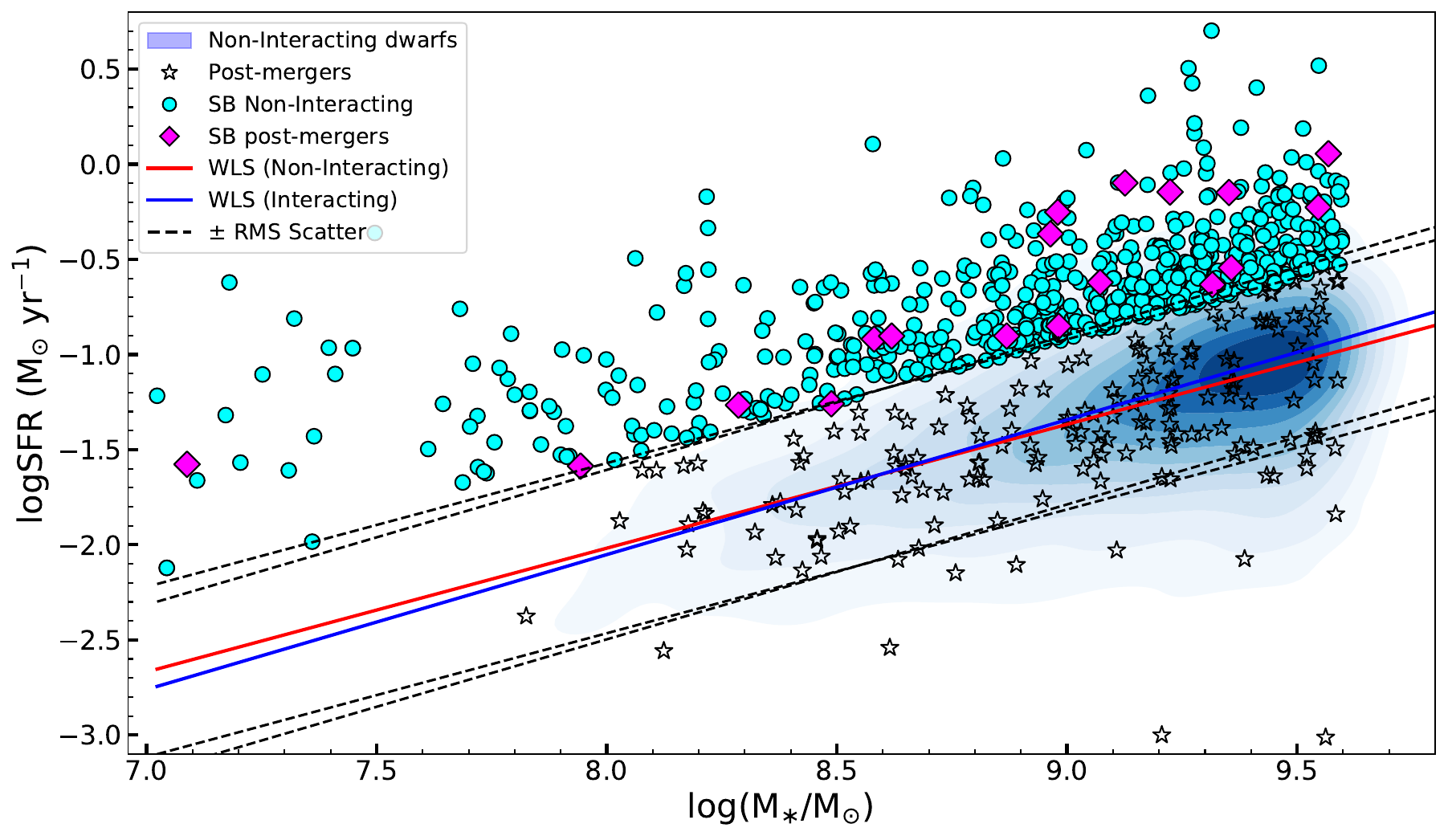}
      \caption{This figure shows the SFMS from Paper I, with SB dwarfs indicated separately above the best-fit + RMS lines: cyan for the non-interacting SB sample and magenta for the post-merger SB sample. The density contours plotted are for non-interacting dwarfs, while individual scatter points denote post-merger dwarfs for the entire sample in Paper I.}
         \label{F1}
   \end{figure}
   
This methodology yields a sample of 656 SB dwarfs, comprising 18 post-merger SB dwarfs and 638 non-interacting SB dwarfs. Figure \ref{F1} shows the SFR vs stellar mass plot in logarithmic scale (similar to Fig.5 of \href{https://www.aanda.org/articles/aa/full_html/2025/10/aa54437-25/aa54437-25.html}{Paper I}), with the identified SB dwarfs shown in cyan (non-interacting SB) and magenta (post-merger SB) points. Dwarf galaxies located within the best-fit $\pm$ RMS lines are classified as main-sequence star-forming dwarfs. This sample of main-sequence star-forming dwarf galaxies (consisting of 154 post-merger dwarfs and 4482 non-interacting dwarfs) is used in subsequent analyses to identify suitable main-sequence control samples, matched in stellar mass and redshift, for the SB samples. \\
\indent As the aim of this study is to compare the distribution of star formation in the two samples of SB galaxies, with respect to the SFMS galaxies, and SFR is directly correlated with stellar mass and redshift, we constructed a sub-sample of non-interacting SB dwarfs matched in stellar mass and redshift for each post-merger SB dwarf. These are selected from the 638 non-interacting SB galaxies within the mass–redshift parameter space of post-merger systems, without replacement, and are iteratively identified as the next-best matches. We aimed to identify up to 15 non-interacting SB controls for each post-merger SB dwarf. To ensure statistical consistency, we performed a Kolmogorov–Smirnov (KS) test between the post-merger and non-interacting control samples, setting the criterion that the KS probability should be greater than 20\% (e.g., \citealt{2008AJ....135.1877E,2011MNRAS.418.2043E,2026MNRAS.552g1698P}). Eventually, we have a sample of 18 post-merger SB dwarf galaxies and a unique control sample of 270 non-interacting SB dwarfs, selected using a KS probability threshold. 
Similarly, for the SB sample of 288 galaxies (18 post-merger SB galaxies and 270 non-interacting SB galaxies), we have corresponding control samples from SFMS galaxies. The control samples are matched in mass-redshift space such that the KS probability is greater than 20\% and there is no duplication of control galaxies. \\
\indent We generated a control sample of 10 main-sequence SF dwarfs for each post-merger SB and 3 main-sequence SF dwarfs for each non-interacting SB galaxy. As our final aim is to compare the spatial variation of star formation in the two samples of SB galaxies with respect to the SFMS, we did not classify the SFMS galaxies as post-merger or non-interacting while selecting the controls. The control sample for both post-merger SB galaxies and non-interacting SB galaxies is chosen from those within the SFMS, including both post-merger and non-interacting SF galaxies. This means that the control sample of a post-merger SB galaxy may include both post-merger and non-interacting SFMS galaxies, matched in mass and redshift. Similarly, the control sample of a non-interacting SB galaxy may contain both post-merger and non-interacting galaxies within the SFMS.
It is to be noted that, in Paper I, we adopted a KS probability threshold of >30\% to construct a control sample of isolated dwarf galaxies (5961) matched to the post-merger dwarf sample (194), following earlier studies of massive galaxies \citealt{2008AJ....135.1877E,2011MNRAS.418.2043E}. In the present study, we initially attempted to retain the same threshold but were unable to obtain sufficient numbers of control galaxies matched in stellar mass and redshift across all subsamples. To ensure that each of the 638 non-interacting SB dwarfs had at least three unique, star-forming control galaxies, we relaxed the threshold to >20\%. For consistency, we adopted this threshold across all subsamples and increased the number of control galaxies for the other subsamples wherever possible.

\indent We use the global SFR, the SFR within the central 3" radius, and the SFR in the outer region (total SFR - SFR within the central 3" radius) for all SB galaxies and their corresponding SFMS control galaxies estimated in \href{https://www.aanda.org/articles/aa/full_html/2025/10/aa54437-25/aa54437-25.html}{Paper I}. To gain further insights into the spatial distribution of star formation, we also estimate and compare SFRs across radial bins within these galaxies. For this, we used the GALEX FUV images of these galaxies and the corresponding background images, downloaded from the MAST portal. 

\section{Analysis} \label{3} 
\begin{figure}[ht!]
   \centering
   \includegraphics[width=9.2cm]{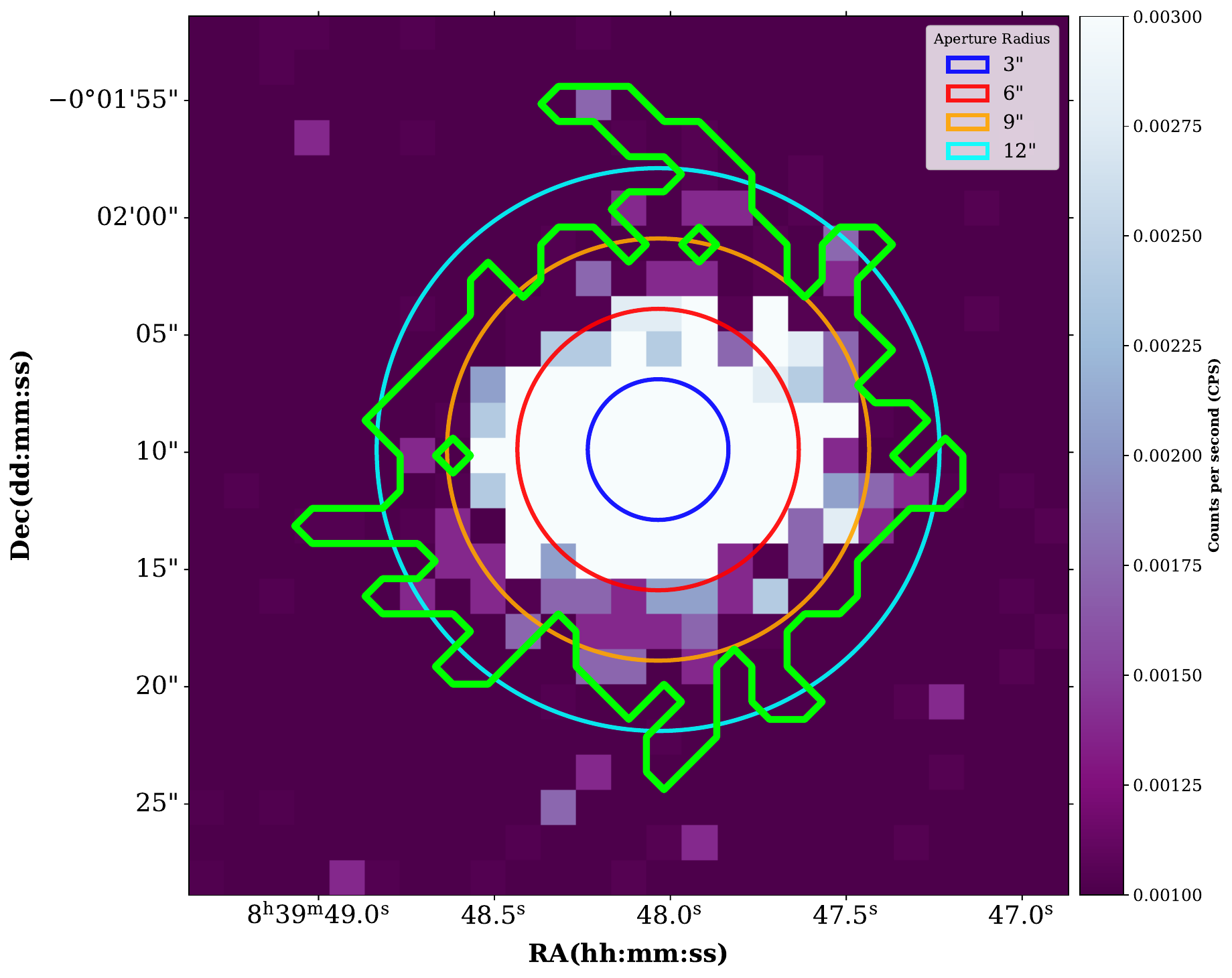}
      \caption{ Galaxy identified by astrodendro is shown with the lime colour structure, and four consecutive circular apertures correspond to different radial bins having radii of 3", 6", 9", and 12" respectively. }
         \label{F2}
   \end{figure}

As explained earlier, we use the SFR and other associated properties of our entire sample from \href{https://www.aanda.org/articles/aa/full_html/2025/10/aa54437-25/aa54437-25.html}{Paper I}. In addition, to effectively analyse star-formation properties across different spatial scales within a sample galaxy, the SFR corresponding to different annular regions is also estimated.
Considering the spatial resolution of the GALEX FUV image of $\sim$ 5.3" for our FUV images, the innermost region is defined as a central circular area with a radius of 3". 
Following the central 3" bin, a series of concentric, annular regions is defined to probe the outer regions of the galaxy. These consecutive radial bins extend outward with increasing radii at 6", 9", and 12" (which extend up to a few tens of kpc at the median redshift of our sample galaxies)
respectively, from the galactic centre, as illustrated in Figure \ref{F2}. 
Based on visual inspection of the FUV images of our sample galaxies, we limited the outermost bin to 12" to ensure detectable FUV flux out to this radius for all galaxies in the sample, including those at the highest redshifts in the sample, while also ensuring that the selected region remains within the structure identified by Astrodendro.\footnote{Our sample galaxies have an irregular shape. So we identified the largest irregular structure in the GALEX FUV image corresponding to the sample galaxy using the Python package Astrodendro \citep{2008ApJ...679.1338R}. For more details, refer to Paper I.}
Furthermore, this size selection helps reduce potential contamination from nearby sources. One post-merger galaxy was excluded from the analysis because the outermost radial bin (12") exhibited contamination from a nearby companion galaxy, as confirmed by visual inspection of archival optical images. Therefore, our final sample consists of 17 post-merger SB dwarf galaxies and their corresponding control samples from non-interacting SB and SFMS galaxies. \\
\indent We adopted the same steps as described in \href{https://www.aanda.org/articles/aa/full_html/2025/10/aa54437-25/aa54437-25.html}{Paper I} to estimate the SFR from the observed FUV images. The FUV flux in each annulus is calculated by converting the counts per second to flux using the calibration relations for the GALEX FUV filter, as given in
\cite{2007ApJS..173..682M}. The observed flux is corrected for foreground extinction using the reddening maps provided by \cite{Schlafly_2011}. Using the distances to our sample galaxies, estimated from their redshifts, we converted the fluxes to luminosities. Finally, the SFR was calculated for each radial bin using FUV luminosities, adopting the calibration from \cite{Murphy_2011}. 
Additionally, to check the effect of PSF smearing on the estimated flux/SFR in different annuli, we took several images having different PSFs and convolved them with slightly broader PSF. We then estimated the SFR in different radial bins for this convolved image. Our estimates were nearly identical to those estimated earlier for unconvolved images, and the changes in SFR were within the error range. The final results, based on comparison of the SFR surface densities across different annular bins of the post-merger and non-interacting samples, did not show any significant change. Therefore, we can conclude that PSF smearing in the outer radial bins does not affect the results of this study. 

\section{Results} \label{4}  
In this study, we compare the estimated SFRs of post-merger SB dwarf galaxies with those of non-interacting samples spanning comparable mass and redshift ranges. Furthermore, a spatially resolved comparison of the SFRs across multiple radial bins within these galaxies was conducted. This spatially resolved comparison is expected to reveal whether the merger-induced starbursts are centrally concentrated, implying gas inflow into the central region, as observed in massive galaxies, or uniformly distributed, suggesting localised tidal compression across the galaxy. 

  \begin{figure*}
  \centering
  \includegraphics[width=9.2cm]{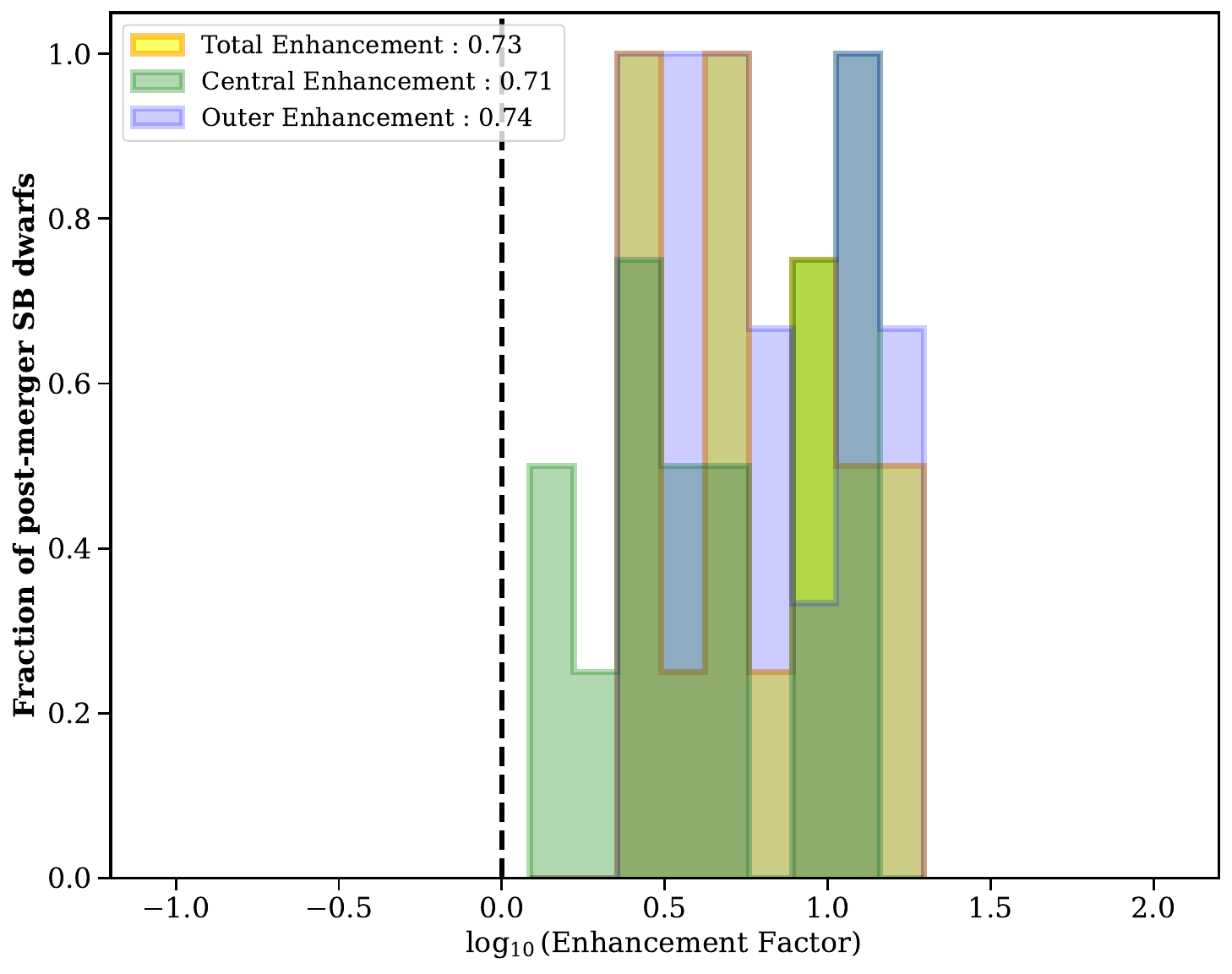}
  \includegraphics[width=9.06cm]{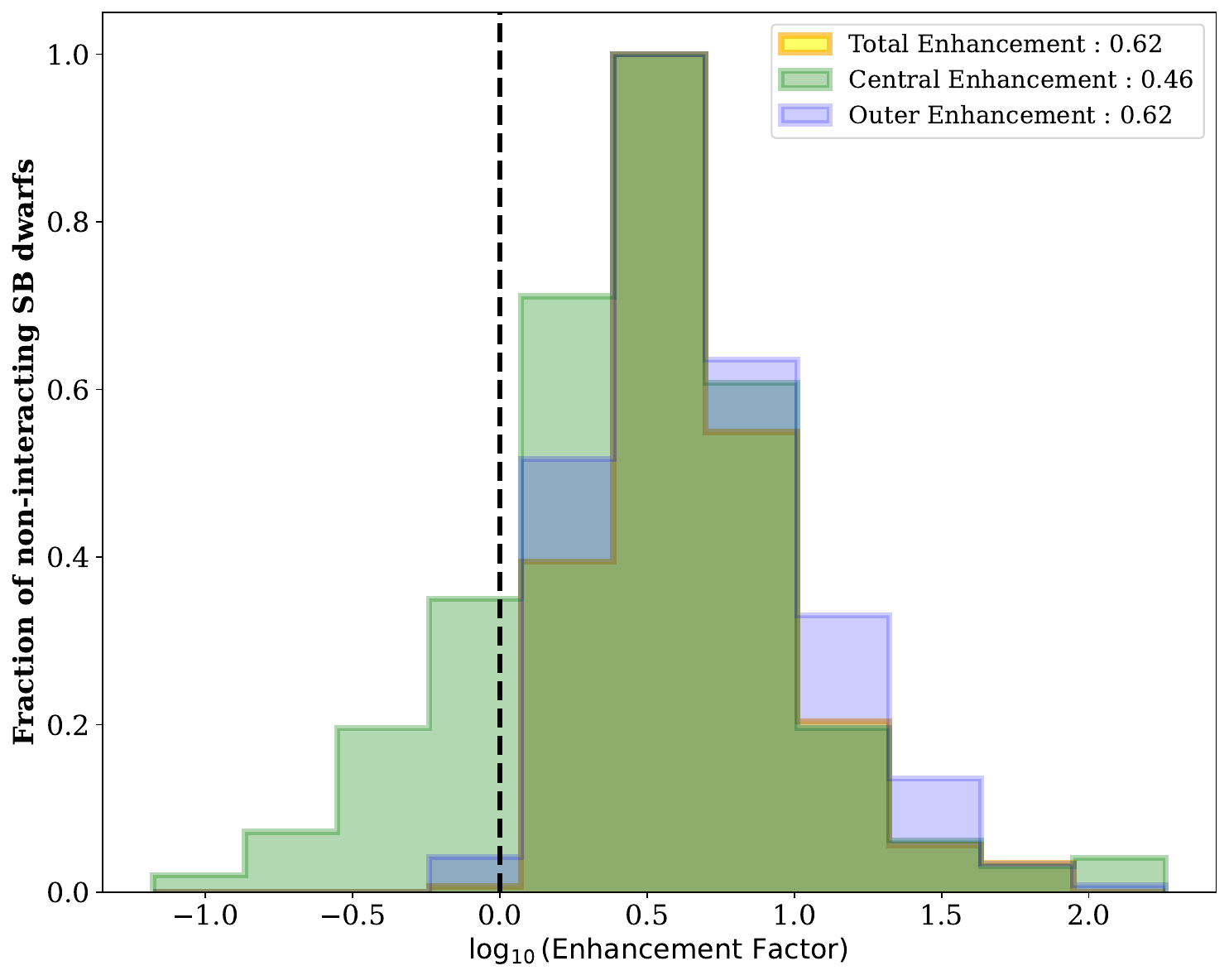}
  \caption{Difference in SFR enhancement of the post-merger and non-interacting systems with respect to their control star-forming dwarfs. Left panel: This figure shows the histogram distribution of SFR offset (log($\Delta$SFR) =  logSFR$_{post-merger~SB}$ $-$  logSFR$_{SFMS~controls}$) for the entire galaxy, central, and outer regions for post-merger compared to their corresponding SF controls. Right panel: This figure shows the similar histogram distribution of SFR offset (log($\Delta$SFR) =  logSFR$_{non-interacting~SB}$ $-$  logSFR$_{SFMS~controls}$) for non-interacting compared to their corresponding SF controls.}
             \label{F4}%
   \end{figure*}

\subsection{Enhancement in star formation} \label{4.1}
To investigate variations in star formation activity, we first compared the SFRs of post-merger SB dwarfs and non-interacting isolated SB dwarfs with respect to their SFMS control dwarfs. This comparison was conducted to determine whether there is a difference in star formation activity, potentially induced by interactions in post-mergers and by secular processes in isolated dwarfs, which might yield divergent SFRs across these two scenarios. 
We calculated the logarithmic ratio of each post-merger SB dwarf's SFR to the median SFR of its matched SFMS control sample. On a logarithmic scale, this ratio is the offset between the log(SFR) of each SB galaxy and the log(Median SFR) of its corresponding control sample. The corresponding SFR offset histograms, illustrated in Figure \ref{F4} (left and right panels), show a positive median offset (log($\Delta$SFR) $>$ 0) in total SFR for both post-merger SB (left panel) and non-interacting SB (right panel) dwarfs with respect to the SFMS dwarfs, suggesting an enhancement in total SFR for SB dwarfs. This is obvious, as the SB galaxies are selected as those above the SFMS + rms in the M$_{*}$ -- SFR plot. However, a notable point is that a similar SFR enhancement is observed in both inner and outer regions of the sample. This is observed in both post-merger SB and non-interacting SB samples. The median offsets in both the central and outer regions are positive and comparable in magnitude. This indicates that, irrespective of the mechanisms that trigger SB in dwarf galaxies, they augment star formation throughout the galaxy. This is especially interesting in the case of post-merger SB dwarfs. It suggests that the central and outer regions of the post-merger are equally enhanced in SFR, rather than being centrally concentrated as observed in massive galaxies.\\
\indent We note that the total, central, and outer SFR enhancements are slightly higher in post-merger SB galaxies than in non-interacting SB galaxies, particularly the central enhancement. The total, central, outer SFRs in post-merger SBs are more enhanced by $\sim$ 1.29 (10$^{0.73}$/10$^{0.62}$ or 0.11 dex), 1.78 (10$^{0.71}$/10$^{0.46}$ or 0.25 dex), and 1.32 (10$^{0.74}$/10$^{0.62}$ or 0.12 dex), respectively, compared to their non-interacting SB dwarfs. This simply indicates that post-merger SB dwarfs have slightly higher SFRs than their non-interacting counterparts, and is consistent with our results in Paper I, which says that, in general, SFR is slightly higher in post-merger dwarfs (irrespective of their location in the SFR vs M$_{*}$ plot) than in their non-interacting counterparts, matched in mass and redshift. But the interesting result here is that the SFR enhancement is uniform across the galaxy in post-merger SB dwarfs. \\

\subsection{Radial variation of SFR density} \label{4.2}
\begin{figure}[ht!]
   \centering
   \includegraphics[width=0.48\textwidth,height=0.36\textwidth]{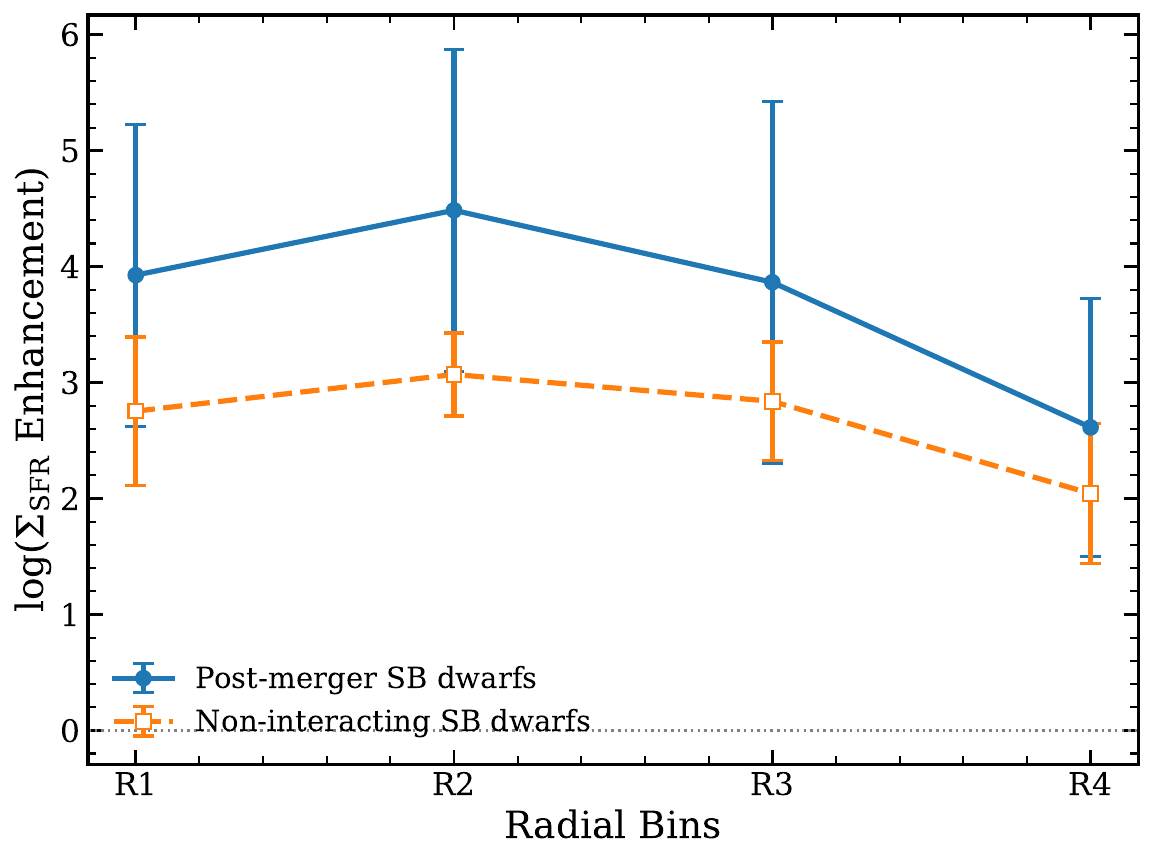}
  \caption {Radial variation of enhancement in SFRD of SB samples relative to the SFMS controls.}
         \label{F5}
   \end{figure}

In the previous sub-section, we find that the SFR of both post-merger and non-interacting SB dwarfs are enhanced in both the inner ($<$3" radius) and outer (outside 3" radius from the centre) regions, and that too, by a similar factor when compared with the corresponding type of SF main sequence dwarfs matched in mass and redshift. This is interesting, as it contrasts with what is observed in massive SB galaxies: post-merger SB galaxies show higher SF in the central regions, while non-interacting SB galaxies show a similar enhancement across the galaxy \citep{2024A&A...690L...4T}. We note that our results obtained in the previous section are based on just two spatial regions (inner and outer 3" radius from the centre), and are also not separated based on physical scales. As the samples are matched in mass and redshift, the physical extent of the galaxies compared is expected to be similar. However, to better understand the spatial variation in star-formation properties in both types of SB galaxies, we compared SFR surface density (SFR per unit area, in kpc$^2$) across 4 radial bins. As described in Section \ref{3}, the four radial bins are at 3", 6", 9", and 12". \\
We estimated the SFRD (SFR normalised with the area of the annulus in kpc$^2$) in 4 annular bins for both sets of SB sample galaxies, and their corresponding SFMS controls. To calculate the offset of the SFRD of SB galaxies, relative to the SFMS controls, we divided each post-merger SB's SFRD by the median SFRD of SFMS  controls across all 4 radial bins. For the post-merger SB sample, we have 17 enhancement values per radial bin, and we have 255 enhancement values per radial bin for the non-interacting SB sample. The median enhancement values in each radial bin and their corresponding errors are plotted in Figure \ref{F5}. The error bars here represent the standard error associated with the median value and is calculated as $$ Error = 1.253\times \frac{\sigma}{\sqrt{N}} $$ where, $\sigma$ is the standard deviation of the SFRD enhancement values in each bin and N is 17 for post-mergers and 255 for non-interacting sample. \\
\indent The plot suggests the following: 1) there is an enhancement in star formation activity across all radial bins for both post-merger and non-interacting SB galaxies, 2) the enhancement is more or less uniform across all the radial bins, for both samples, and 3) indicates that the enhancement factor for the post-merger SB is slightly higher ($\sim$ 1.4 -- 1.5 times in the first three bins) compared to the non-interacting SB, but they are within error bars. These results are consistent with the observations in Section 4.1 that star formation activity is uniformly enhanced in post-merger SB dwarf galaxies rather than centrally concentrated. We note a drop in the enhancement factor in the last radial bin, especially for the post-merger SB sample. This might be due to the irregular morphology of dwarf galaxies, especially the disturbed morphology of post-merger dwarf galaxies, in which an azimuthally averaged SFRD estimate for the last radial bin may not be appropriate. \\

\section{Discussion} \label{5}
Despite numerous and extensive investigations, the mechanisms that trigger, fuel, and terminate starburst activities in galaxies remain poorly understood. High-mass starburst galaxies are typically associated with dramatic external events, such as major mergers and tidal interactions (e.g., \citealt{1996ARA&A..34..749S}), suggesting that external, violent events are necessary to trigger intense star formation. In contrast, starbursts in isolated spiral galaxies appear to be associated with stellar bars \citep{1998ARA&A..36..189K,2004ARA&A..42..603K}, suggesting that SB events may result from secular processes such as internal instabilities, bar-driven gas inflow, and stochastic gas dynamics. However, recent studies \citep{2024A&A...690L...4T} suggest that merger-driven and secular-driven SB galaxies in the high-mass regime exhibit distinct observable characteristics in the spatial distribution of star formation. In merger-driven SB galaxies, star formation is enhanced and concentrated centrally, whereas in secular-driven SB galaxies, it is uniformly enhanced throughout the system. \\
\indent The situation becomes more complex for low-mass SB galaxies, such as blue compact dwarfs (BCDs; \citealt{2003ApJS..147...29G}) and HII galaxies (e.g., \cite{1996ApJS..102..189T}), which typically lack stellar bars and spiral arms. Several studies argue that external mechanisms are crucial in triggering starbursts even in these low-mass systems, including minor mergers, close-range tidal interactions with nearby low-mass companions, and the accretion of primordial or metal-poor gas from the intergalactic medium (e.g. \citealt{2010MNRAS.406.1238E,2010A&A...517A..85L,2014A&A...566A..71L}). If starbursts are the result of external perturbations, then one would expect to see distinct kinematic and morphological signatures of interaction in the gaseous disk of the galaxy. However, a significant number of observed SB galaxies are isolated, with no clear evidence of recent or ongoing external tidal interactions. The existence of these isolated SB systems strongly suggests that potential starburst trigger mechanisms are not solely dependent on external forces but might involve internal or secular drivers. In this case, whether SB galaxies, driven by different mechanisms, also exhibit distinct observable signatures in the low-mass regime remains poorly understood. 

In this study, we compare the spatial distribution of star formation activity in post-merger SB dwarfs with that of non-merging SB control dwarfs across different mass and redshift bins in order to disentangle the effect of merger-driven and secular SB events in the low-mass regime. We did not find a clear distinction in the spatial distribution of star formation in post-mergers relative to non-interacting starburst dwarfs. A similar enhancement in SFR was observed in the inner and outer regions of post-merger SB dwarfs, in a similar manner as in the non-interacting SB sample dwarfs. Further, we investigated the radial variation of SFRD for post-merger and non-interacting dwarfs. We find that there is an enhancement in star formation activity across all radial bins for both post-merger and non-interacting SB galaxies, and the enhancement is more or less uniform across all the radial bins. This indicates that star formation in low-mass SB galaxies is uniformly enhanced throughout the galaxy, regardless of their SB triggering mechanism. We also found that the enhancement factor is slightly higher for the post-merger SB dwarfs, though the values are within error bars for both samples. \\
\indent The observed lack of central concentration of star formation in our sample of post-merger SB dwarfs is in contrast with what is observed in massive merger systems where star formation is centrally concentrated \citep{2014ApJ...789L..16L,2015ApJ...807L..16K,2020MNRAS.492.2075B,2022MNRAS.516.4922R}. In our sample of post-merger SB dwarf galaxies, we observed more uniformly enhanced star formation across extended regions. This is similar to what is theoretically predicted in models and simulations  (\citealt{1992ApJ...387..152J,2020ApJ...891....2L}) of low-mass SB galaxies, and is observed in individual systems \citep{Privon_2017}). A recent study by \cite{2024MNRAS.533.3771L} also suggested that interacting dwarfs display comparatively bluer colour profiles across all radii relative to their non-interacting counterparts, indicating an enhanced star formation throughout the entire galaxy rather than restricting it to the central region. These results are well aligned with the conclusion of our study. Although this study uses the (g-i) colour profiles to examine the radial variation of star formation. Optical colours are sensitive not only to the recent star-formation activity but also to metallicity and the relative contribution of different stellar populations. They are subject to the age--metallicity degeneracy, whereby changes in the age and metallicity of a stellar population can produce similar changes in its observed colour \citep{1994ApJS...95..107W,2009MNRAS.397..695C}. In addition, observational detection limits and surface-brightness effects can introduce biases in the inferred colour profiles in the outskirts of LSB dwarfs. 
Therefore, a bluer (g-i) colour does not necessarily signify higher star formation activity, but it could be a consequence of differences in the underlying stellar population. However, in our study, we use FUV, which is a direct indicator of recent star formation. Hence, it is a more robust criterion that suggests results similar to those obtained by \citep{2024MNRAS.533.3771L}.
Our observed result may arise from shallower gravitational potential wells of low-mass galaxies, which allow merger-driven gravitational perturbations, gas inflows, and feedback-induced motions to redistribute the interstellar gas during these interactions or merger events \citep{2022A&A...661A.136G,2024A&A...690A.286A}. Low mass galaxies are also known to have extended gas envelopes \citep{2016MNRAS.459.1827P}, and during mergers, compression of these extended gas envelopes leads to clumpy star formation over extended regions, and not centrally concentrated \citep{Privon_2017}. \\
\indent While saying that star formation is not centrally concentrated in post-merger SB dwarfs, we note that accurately determining the dynamical centre in dwarf irregular systems is challenging \citep{2001AJ....122.3070O,2024MNRAS.527.5339B}. They often lack the clear, centrally concentrated regions like bulge as found in massive spirals, and their gas and stellar distributions can be highly asymmetric \citep{2018ApJ...855....7H}. Consequently, uncertainties in determining the galaxy centre may result in the misclassification of regions as central. As a result, the SFR estimated for these regions may not reflect the actual SFR of the central regions. However, we do not find an SFR/SFRD enhancement systematically offset from the region we adopted as the galaxy centre. Instead, the enhancement is distributed relatively uniformly across the extended regions of our sample. Thus, the uncertainties in identifying the precise location of the dynamical centre are unlikely to affect our results.\\ 
\indent Another caveat is related to the difficulties involved in detecting fainter morphological signatures of merger remnants in lower-mass dwarfs. Therefore, if merger signatures are missed in these fainter, low-mass systems, some of the post-merger SB galaxies may be misclassified as non-interacting controls. However, even if a considerable fraction of our non-interacting SB control samples were indeed post-merger SB galaxies, the median values of non-interacting SB galaxies also show a uniformly enhanced star formation across the extended regions. Thus, our results are not likely to be affected by the misclassification of some post-merger SB dwarfs as non-interacting SB dwarfs. \\
\indent Our overall results suggest that, within dwarf galaxies, the triggering mechanism may not be a dominant factor in determining the nature and extent of star formation. The observational evidence from the selected sample does not indicate a clear dichotomy; instead, it points towards broadly similar star-formation properties across the different scenarios considered, with slight variations within the uncertainties. 
In future, deep, high-resolution observations of larger samples of dwarf galaxies would be useful for better understanding the spatial distribution of star formation in low-mass galaxies.

\section{Summary} \label{6} 
Starburst episodes within a galaxy can be instigated by various mechanisms, encompassing both internal and external evolutionary pathways. Although galactic interactions represent the most significant trigger for starbursts, they are not the sole responsible mechanism. For instance, vigorous star formation can be initiated by inward gas flows in barred galaxies, or by spontaneous global gravitational instability in dwarf galaxies. It is clear that starbursts, although relatively short-lived, form an important phase in galaxy evolution. In addition to their crucial role in galaxy evolution, starbursts are also very useful for investigating the process of star formation itself. 
However, whether changes in spatial distribution triggered by external processes, such as mergers, differ intrinsically from those induced by internal, or secular, starbursts remains poorly understood, especially in low-mass galaxies. \\
\indent In this study, we investigated the star formation properties of a sample of 272 isolated starburst (i.e., with no massive galaxy, M$_{*}$ $>$ 10$^{10}$ M$_{\odot}$, within a 1 Mpc$^3$ volume) dwarf galaxies consisting of 17 post-merger and 255 non-interacting galaxies, matched in mass and redshift. They span a stellar mass range of $ 10^{7} - 10^{9.6} M_{\odot} $ and a redshift range of 0.01 -- 0.12. We use the FUV imaging data from the GALEX mission to estimate the SFRs of our sample galaxies. We did not find a clear distinction in the star formation triggered in post-mergers relative to non-interacting starburst dwarfs. A similar enhancement in SFR, both in the inner and outer regions, was observed in both post-merger and non-interacting SB sample dwarfs relative to their respective SFMS controls matched in mass-redshift space. Furthermore, we investigated the spatial variation of starburst activity by comparing SFR surface density (SFRD) across radial bins for post-merger and non-interacting dwarfs. We find that there is an enhancement in star formation activity across all radial bins for both post-merger and non-interacting SB galaxies, and the enhancement is more or less uniform across all the radial bins, and within error bars for both samples. This indicates that star formation in low-mass SB galaxies is uniformly enhanced throughout the galaxy, regardless of their SB triggering mechanism. This contrasts with what is observed in massive galaxies, where star formation is centrally concentrated in the merger-driven SB and uniformly distributed in the secularly driven SB. Deeper and higher-spatial-resolution studies can provide greater insight into the spatial distribution of star formation in low-mass galaxies. \\

\begin{acknowledgements}
      This publication uses archival data from GALEX (Galaxy Evolution Explorer), a NASA Small Explorer mission. RC acknowledges financial support from the Council of Scientific and Industrial Research (09/0890(17755)/2024-EMR-I). SS acknowledges the support from the Science and Engineering Research Board of India through the POWER grant (SPG/2021/002672) and support from the Alexander von Humboldt Foundation.
      This work utilised the community-developed core Python packages for Astronomy, like Astrodendro {\url{(https://dendrograms.readthedocs.io/en/stable/)}}, 
      Astropy \citep{2013ascl.soft04002G,Price-Whelan_2018,2022ApJ...935..167A},
      Matplotlib \citep{2007CSE.....9...90H}, Photutils \citep{2024zndo..10967176B}, SciPy \citep{2020NatMe..17..261V} and Numpy \citep{2020Natur.585..357H} software packages to analyze and process astronomical data.
\end{acknowledgements}

\bibliographystyle{aa} 
\bibliography{reference} 

\end{document}